\documentclass[a4paper]{jpconf}
\usepackage{graphicx}
\begin{document}
\title{The origin of ultra high energy cosmic rays}

\author{Pasquale Blasi}

\address{INAF/Osservatorio Astrofisico di Arcetri, Largo E. Fermi, 5
  50125 Firenze (Italy)}

\ead{blasi@arcetri.astro.it}

\begin{abstract}
We briefly discuss some open problems and recent developments in
the investigation of the origin and propagation of ultra high energy 
cosmic rays (UHECRs). 
\end{abstract}

\section{Introduction}

The detection of cosmic rays with energy in excess of $10^{20}$ eV
puzzles the scientific community, mainly for two reasons: 1) most
acceleration mechanisms need to be stretched to their extreme to allow
for the generation of particles with this energy; 2) Energy losses of
protons with energy high enough to suffer photopion production are
such that, if the source distribution is homogeneous on scales larger than 
the loss length, a flux suppression should be expected in the
spectrum, the so-called {\it GZK feature} \cite{gzk}. It is not clear
yet if this feature  has been observed.

The importance of these two issues is here addressed in the context of
the most recent findings in the investigation of particle acceleration
at shock fronts, and in the light of the up-to-date measurements of
the cosmic ray flux, anisotropy and chemical composition. A more
extended discussion of these topics and their connection to the
physics of lower energy cosmic rays can be found in two recent reviews 
\cite{hillas,blasirev}.

This review is organized as follows: we discuss some developments in
the theory of particle acceleration at shock waves in
Sec. \ref{sec:shocks}). In Sec. \ref{sec:transition} we address the
important issue of identifying the energy region in which the
transition from a galactic to an extragalactic origin of cosmic rays
takes place. In Sec. \ref{sec:gzk} we discuss the current measurements
of the spectrum and anisotropies of UHECRs. We conclude in Sec. 
\ref{sec:summary}.

\section{Particle Acceleration at shock fronts}
\label{sec:shocks}

Diffusive particle acceleration at collisionless shock waves
(hereafter Diffusive Shock Acceleration or DSA) remains
one of the most promising acceleration mechanisms to energize charged
particles to very high energies. Historically, the theory of DSA was
first developed for the case of non relativistic shock fronts 
\cite{kry,bo,bell78} and later for the case of relativistic shocks
\cite{peacock81}. A full theory which describes particle acceleration
at shocks of arbitrary speed and arbitrary scattering properties of 
the background medium was recently formulated \cite{vietri,bvietri}
and applied to a variety of cases \cite{bvietri,morlino}.
All these approaches are based on the assumption that the accelerated 
particles exert no dynamical reaction on the shock, the so-called 
{\it test particle} approximation ($TPA$). 

Recent developments in this field are aimed at addressing the
following issues: 1) Is the {\it test particle} approximation broken 
and if so, what are the implications? 2) What provides the scattering 
agents that allow particles to return to the shock front from the 
upstream fluid? This question, as discussed below, is intrinsically 
related to the determination of the maximum energy of the accelerated 
particles. 

The two questions above, as we will find, are not independent from 
each other. Regrettably, most investigations trying to address these
issues have been carried out for the case of non relativistic shocks,
with a few exceptions. 

The {\it test particle} approximation was first relaxed in the
context of thermodynamical models, known as {\it two-fluid} models
\cite{two}, in which the accelerated particles and the 
thermal plasma are described as two separate fluids interacting 
through the pressure exerted by the accelerated particles on the 
background shocked plasma. This class of models does not provide
informations on the spectrum of the accelerated particles, but only on
their general thermodynamical properties and their effect on the 
background plasma. The first attempts to
gather information on the spectrum were carried out in
\cite{bland,eichler}. These models led to the modern {\it kinetic models}
\cite{malkov1,malkov2,blasi1,blasi2,vannoni,elena} which provide also
the spectrum of the accelerated particles. In this sense, kinetic
models include two-fluid models. At the same time, several numerical
approaches were developed \cite{numerical}. A relatively recent
up-to-date technical review of these subjects can be found in 
\cite{maldru2001}. 

It is qualitatively easy to explain why the breaking of the $TPA$ may
play a very important role for particle acceleration: the typical
spectrum of accelerated particles at a strong shock (very large 
Mach number) within the TPA is $f(p)\propto p^{-4}$, which 
contains a divergent amount of energy when integrated over momentum, unless a
high momentum cutoff, $p_{max}$ is imposed. Since $p_{max}$ is
determined by environmental parameters (e.g. magnetic field, size or 
age of the accelerator), even with a given $p_{max}$ and at fixed
number of particles accelerated to suprathermal energies, it may
happen that the pressure in the form of accelerated particles, $P_{CR}=
\int_0^{p_{max}} dp 4\pi p^3 v(p) f(p)$, exceeds the total 
kinetic pressure $\rho u^2$ crossing the shock front. This is clearly 
unphysical and suggests that the reaction of the particles on the 
shock may become important. Moving towards the shock front from
upstream infinity, the pressure of the accelerated particles
increases, and the fluid is slowed down to form a {\it shock
  precursor}. The reaction of the accelerated particles implies that
the total compression factor between upstream infinity and downstream
can vastly exceed 4, while the compression factor at the subshock
remains bound to be $\leq 4$. The kinetic models have convincingly
demonstrated that this situation corresponds to non power-law concave spectra
of the accelerated particles: the spectrum is harder than $p^{-4}$ at
low momenta and harder than $p^{-4}$ for high momenta. As a
consequence, most pressure is exerted by the highest energy particles
and this implies that the few particles with $p\sim p_{max}$ that can
escape to upstream infinity carry enough energy to make the shock
radiative and thereby more compressible, namely the spectrum tends
to get harder at high energies. Moreover, if a sufficiently large
amount of energy is accumulated in the accelerated particles, the
adiabatic index of the fluid tends to $4/3$ instead of $5/3$, and
again this makes the shock more compressible. This set of phenomena
all point in the direction of contributing to the shock modification 
and to an enhancement of the acceleration efficiency. In the absence
of Alfv\`en heating or other processes that heat up the upstream fluid
in a non-adiabatic way, very high efficiencies of acceleration can be
achieved (e.g. \cite{elena}). 

All the calculations developed so far assume that the diffusion
coefficient that determines the efficiency of the scattering of
particles in the upstream and downstream plasmas is an input to the
problem. It is well known that at least for the case of
supernova remnants in the interstellar medium (ISM), the level of
magnetic turbulence is insufficient to allow for energization of the
particles to energies of astrophysical interest. The general picture
of DSA admits that local magnetic scattering is self-generated in the
shock vicinity due to streaming instability \cite{bell78}. The amount
of turbulence that can be generated in this way determines the maximum
momentum of the accelerated particles (see \cite{lc83} for the case
of supernova remnants). If stationarity of the acceleration process
and of the generation of Alfv\`en waves is assumed, the maximum level
of turbulence can be shown to be \cite{vmc82}:
\begin{equation}
\frac{\delta B^2}{B^2} = 2 M_A \frac{P_{CR}}{\rho u^2},
\label{eq:saturation}
\end{equation}
where $M_A$ is the Alfv\`en Mach number. If $P_{CR}\sim \rho u^2$,
the amplification of the magnetic field with respect to the background
undisturbed field can be as high as $\sim M_A^{1/2}$. Several
processes can reduce the effective magnetic field to values much lower 
than those found through Eq. \ref{eq:saturation} \cite{ptu}. On the
other hand, Eq. \ref{eq:saturation} immediately shows that the linear
theory that it is based upon is easily broken in the description of 
efficient particle acceleration at shocks (namely if $P_{CR}\sim \rho u^2$).

In a series of recent investigations \cite{bl00,bell2004} the
possibility was advanced that the saturation level of the turbulent
magnetic field may be even higher than what found through
Eq. \ref{eq:saturation}. In \cite{bell2004}, Bell discussed in detail
the generation and growth of non-resonant, non-alfvenic waves induced
by cosmic rays in the shock vicinity: in the reference frame of the
upstream plasma, the shock moves toward the observer and carries a
current of positively charged particles $\vec J_{CR}$(the accelerated
particles). Due to the condition of charge neutrality, a return
current $\vec J_{ret}$ is established in the upstream plasma. 
In the frame of the upstream plasma, the plasma itself is subject to 
a force $\vec J_{ret} \times \vec B$. This force vanishes at the zero
order in the case of a parallel shock, but after taking into account
the perturbations of the equation of motion and of Maxwell equations,
this term remains in the form of a forcing term. The dispersion
relation of the waves can be shown to develop a branch of non-alfvenic
modes with a dominant imaginary part of the frequency, which are
therefore almost purely growing. In \cite{bell2004} the non-linear
part of the evolution of the perturbations was followed through a
hybrid numerical simulation. The author of Ref. \cite{bell2004}
reaches the conclusion that the saturation level of the unstable 
modes is well described by
\begin{equation}
\frac{\delta B^2}{B^2} = M_A^2 \frac{u}{c} 
\frac{P_{CR}}{\rho u^2}.
\label{eq:sat_bell}
\end{equation}
Whether the saturation level is correctly described by this expression
or is rather determined by one of the several effects that could not
be included in the calculations of \cite{bell2004} is still to be
seen. However, if Eq. \ref{eq:sat_bell} is in fact correct, then the
magnetic field can be amplified in the proximity of a shock
to very large values $\delta B/B\sim M_A (u/c)$, if $P_{CR}\sim \rho
u^2$, thereby opening the possibility that acceleration to very high
energies may be achieved. 

The implications of this result for galactic cosmic rays have been
discussed in \cite{hillas,blasirev}. It is easy to extend those
conclusions to astrophysical scenarios where DSA is at work and can
possibly lead to the generation of UHECRs, though at present no
systematic investigation has been carried out. Due to the many physical
processes involved (generation, growth and damping of non-alfvenic
perturbations) the role of this mechanism of magnetic field
amplification should be studied on the case-by-case basis. 

\section{The transition from galactic to extragalactic origin} 
\label{sec:transition}

The first step towards unveiling the origin of UHECRs is to
understand at which energies cosmic rays become of extragalactic
origin. The transition from galactic to extragalactic cosmic rays 
has historically been identified with the ankle and 
interpreted as the intersection of a rapidly falling galactic 
spectral component and a flat spectrum of extragalactic cosmic 
rays. In \cite{bere1,bere2} it was pointed out that the combination 
of pair production energy losses and adiabatic energy losses due 
to the expansion of the universe generates a second knee and a 
dip in the spectrum of extragalactic cosmic rays, consistent with the
observations (see Fig. 6 for predictions obtained following the 
calculations of \cite{beregrigo,bere1,bere2}).

\begin{figure}[h]
\begin{center}
\includegraphics[width=18pc]{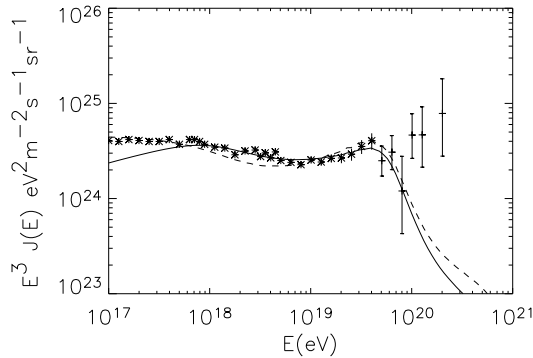}\hspace{2pc}%
\end{center}
\caption
{\label{fig:27vs24} Spectrum of cosmic rays of extragalactic origin for an
  injection spectrum $\propto E^{-2.7}$ and no luminosity evolution 
  (solid line) and for injection spectrum $\propto E^{-2.4}$ and
  luminosity evolution $L(z)\propto (1+z)^4$ (dashed line). The data 
  are from Akeno and AGASA.  }
\end{figure}

In this model the transition from galactic to extragalactic cosmic
rays takes place at energies below $\sim 10^{18}$ eV, where the 
galactic cosmic ray component would be cut off. In its basic form, this
scenario implies that cosmic rays are injected at extragalactic
sources with an injection spectrum $E^{-2.7}$ and no luminosity
evolution (solid line in Fig. \ref{fig:27vs24}). Although it is 
a worse fit, also the case of sources with luminosity evolving with
redshift as $L(z)\propto (1+z)^4$ and with injection spectrum $E^{-2.4}$ 
shows a dip-like feature (dashed line in Fig. \ref{fig:27vs24}). 
In both cases, even small levels of magnetization of the intergalactic 
medium would induce a low energy suppression of the flux, which might 
in fact improve the fit to the all-particle spectrum \cite{lem,alo}. 
The contributions of many sources with different values of the maximum 
energy $E_{max}$ was shown in \cite{michael} to mimick a steeper 
injection spectrum, though requiring much less energy injection as 
compared with the case of an injection spectrum $E^{-2.7}$.

A clear prediction of this model is that the Galactic component of
cosmic rays should be vanishing at energies $\leq 10^{18}$ eV. The 
KASCADE data \cite{kascade} show that above the knee the chemical 
composition of cosmic rays becomes increasingly more dominated by 
heavy elements. 
Moreover, these data suggest that rigidity dependent knees in the
single chemical components appear, and that cosmic rays in the 
energy region around $10^{17}-10^{18}$ eV are expected to be mainly
iron nuclei. At these energies the flux is also expected to vanish, 
if the rigidity dependent knees observed for protons and helium nuclei 
\cite{blasirev} are assumed to be present for heavier elements as
well. This observational scenario appears to provide circumstantial 
evidence that the spectrum of galactic cosmic rays may be terminated 
at $10^{17}-10^{18}$ eV. 

A more serious concern for the model of \cite{bere1,bere2} is related 
to the chemical composition: the dip is in fact likely to disappear if 
a small contamination (with solar-like abundance) of nuclei heavier than
hydrogen is present at the source \cite{bere2,angela}. Magnetic
horizon effects related with nuclei with different charge to mass
ratio might somewhat mitigate the relevance of this issue \cite{sigl}. 
An accurate measurement of the chemical composition of cosmic rays in
the transition region appears to be of crucial importance. 

\section{Spectrum and small scale anisotropies of UHECRs}
\label{sec:gzk}

UHECRs present us with at least two questions. The first one is 
related to the search for an acceleration process and an 
acceleration site at which these particles win against energy 
losses or finite size of the accelerator and succeed in becoming 
UHECRs. The second is related to their propagation: for sources 
distributed homogeneously in the universe, particles for which 
the reaction of photopion production is kinematically allowed 
suffer rapid energy losses, with a loss length of $\sim 20$ Mpc 
at $10^{20}$ eV. This results in a suppression
of the flux at the Earth, the so-called GZK feature \cite{gzk}. The shape
of this feature is extremely model dependent: local overdensities of
sources, the injection spectrum, local magnetic fields and source
evolution all play a role in changing the shape of the GZK feature. 

In this section we deal with the second point, namely the detection of
the GZK feature and the role that small scale anisotropies in the
directions of arrival can play in unveiling the sources of UHECRs. 

\begin{figure}[h]
\includegraphics[width=18pc]{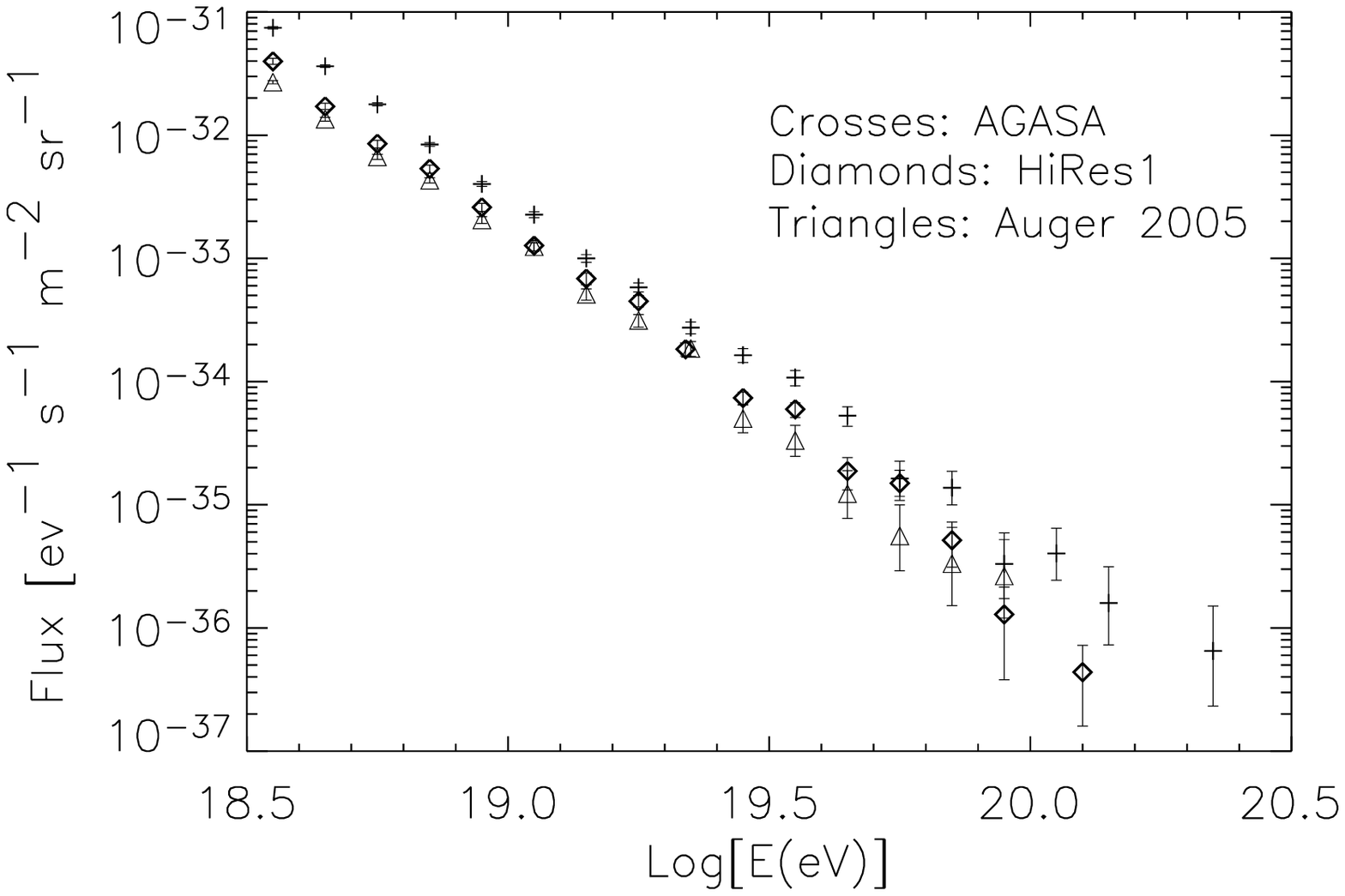}\hspace{2pc}%
\includegraphics[width=18pc]{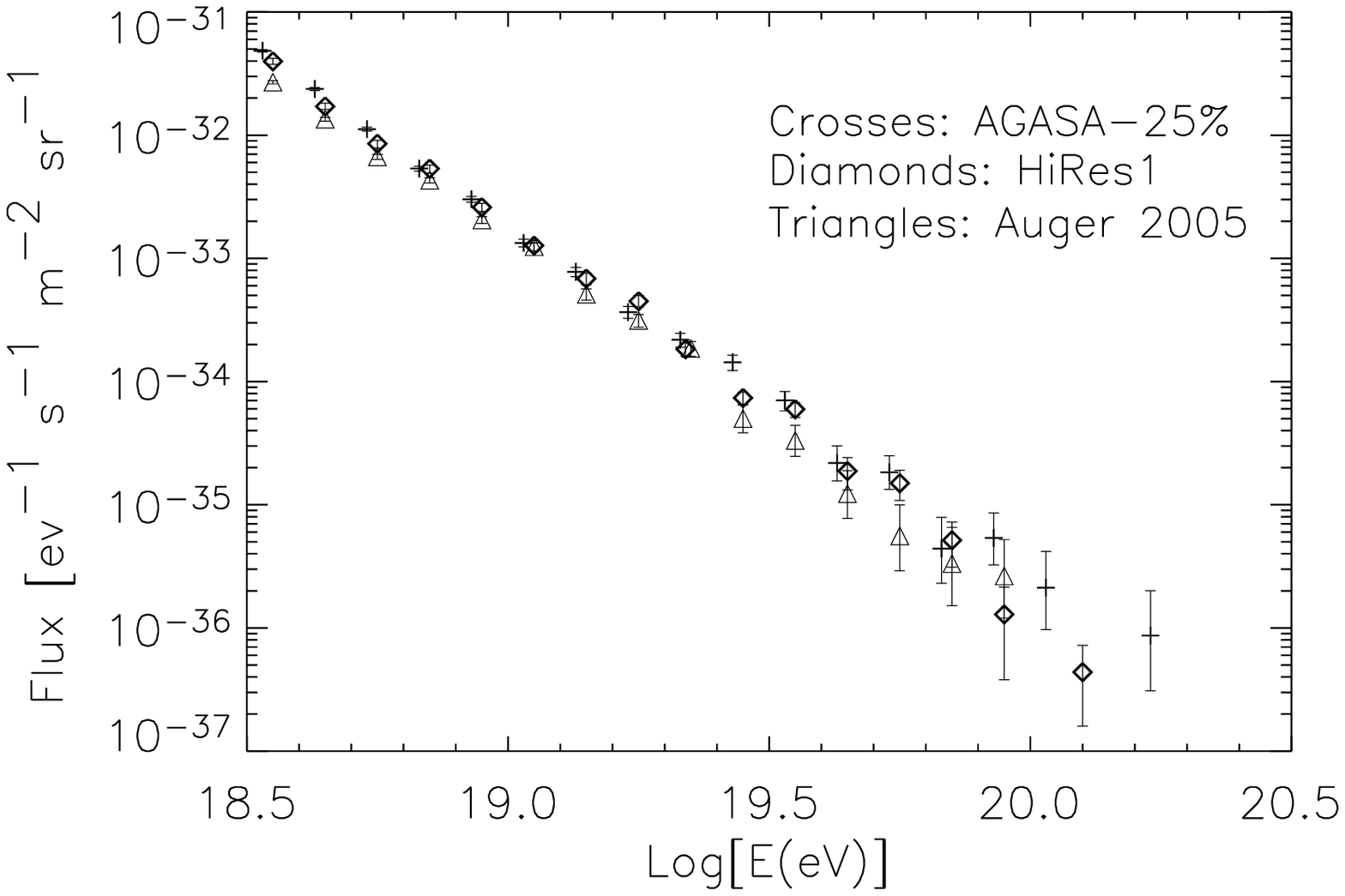}\hspace{2pc}%
\caption
{{\it Left Panel)} Spectrum of AGASA, HiRes and Auger. 
{\it Right Panel)} Spectrum of AGASA, HiRes and Auger with a
correction of the AGASA energy for a possible systematic error in the 
energy determination by $-25\%$.}
\label{fig:exp}
\end{figure}

The very low fluxes expected at energies around and above $10^{20}$ 
eV makes the detection of the GZK feature very problematic. 
Experiments that have operated so far have not been successful 
in either detecting the GZK feature or proving its absence with a
sufficiently high statistical significance. The spectra of AGASA, 
HiRes and the newly released data from the Pierre Auger telescope 
\cite{augerdata} are plotted in Fig. \ref{fig:exp} (left panel).
In the right panel of the same figure we show the spectra when the 
AGASA energies have been shifted downwards by $25\%$. 

It is rather remarkable that all experiments are in 
substantial agreement, and that the few differences can be 
interpreted in terms of rather small systematic energy shifts in one
or more of the experiments. Several Monte Carlo simulations (see
\cite{dbo1,dbo2}) have convincingly showed that the differences in
the high energy parts of these spectra have a rather low statistical 
significance, of the order of $2-3\sigma$. This means that for a given
statistics of events above $10^{19}$ eV, the number of events with 
energy above $10^{20}$ eV is low enough that is affected by
statistical fluctuations. 

\begin{figure}[h]
\begin{center}
\includegraphics[width=19pc]{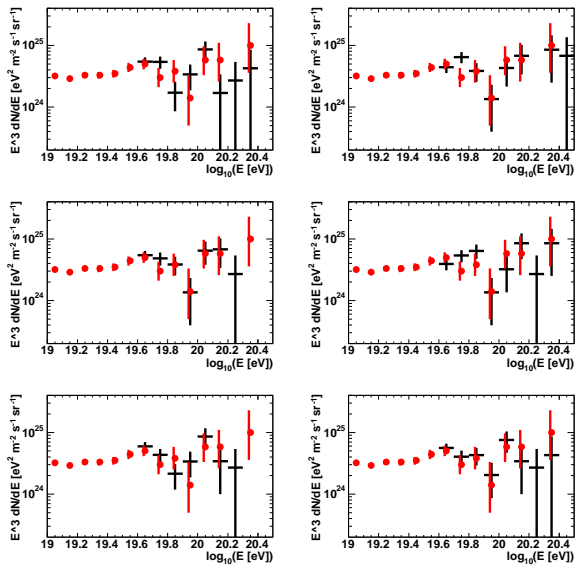}\hspace{2pc}%
\end{center}
\caption
{Some of the simulated realizations with number of events
  equal to or larger than the actual number detected by AGASA \cite{dbo2}.}
\label{fig:agasalike}
\end{figure}

These statistical considerations are the result of averages over large
samples of realizations of source distributions, therefore one might
wonder whether the individual spectra of those realizations which have a
large number of events at ultra high energies are similar to the AGASA 
spectrum or not. In Fig. \ref{fig:agasalike} (from \cite{dbo2}) we show
the spectra of UHECRs obtained in some of the realizations that showed 
11 or more events above $10^{20}\rm eV$ (the error bars here are just 
poissonian, namely they have the same meaning as in
Fig. \ref{fig:exp}). These spectra closely resemble the AGASA 
spectrum, namely they do not show any GZK suppression, despite 
the fact that in the lower energy region they all fit the data 
quite well. This shows that an AGASA-like spectrum is not that 
improbable, even if the {\it average} cosmic ray spectrum can 
be expected to show a GZK feature \cite{dbo2}.

The spectrum of cosmic rays by itself does not contain enough 
information to determine the type of sources of UHECRs.
A first hint at the type of sources can come from the
identification of small clusters of events with arrival directions
which differ little. Such a signal of small scale
anisotropies (SSA) was first claimed by the AGASA collaboration 
\cite{agasa_ssa}. The signal was however shown to have a 
low statistical significance \cite{finley}. No evidence of SSA 
has been found so far in the HiRes data \cite{hires_ssa}. 

On the other hand, if astrophysical point sources are believed to accelerate
UHECRs, then SSA are to be expected. It was shown in \cite{danny} that, if
the AGASA results were confirmed, the number density of sources could be
estimated to be around $10^{-5}\rm Mpc^{-3}$, with a quite large
uncertainty due to the very limited statistics of events available
(see \cite{allssa} for other estimates). In \cite{dbo2} the authors
show that the spectrum of AGASA appears to be not fully consistent
with the detection of the SSA by the same experiment. The possibility
of coupling the information on the spectrum and SSA to gather better 
information on the number of sources will probably prove useful with
the upcoming data from the Pierre Auger Collaboration. 

\section{Summary}
\label{sec:summary}

We briefly discussed three issues related to the origin and
propagation of ultra-high energy cosmic rays. In particular we
summarized some recent developments in the investigation of particle
acceleration at non-relativistic shock fronts, including the non
linear dynamical reaction of the accelerated particles on the shock
and the generation of waves by the accelerated particles. We also
addressed the important issue of the transition from galactic to
extragalactic origin of the observed cosmic rays. Finally we discussed
the status of the measurements of spectrum and anisotropies of
UHECRs.

\end{document}